\documentclass[letterpaper,twoside,english]{article}
\usepackage[T1]{fontenc}
\usepackage[latin1]{inputenc}
\usepackage[authoryear]{natbib}

\makeatletter

\usepackage{setspace}
\usepackage{graphicx}
\usepackage{latexsym,times}
\usepackage[intlimits]{amsmath}
\usepackage{amsfonts,amssymb}
\DeclareSymbolFontAlphabet{\mathbb}{AMSb}
\usepackage[mathscr]{eucal}
\usepackage[textstyle,cdot,squaren,thickspace,thickqspace]{SIunits}
\usepackage{fancyvrb}
\usepackage{url}
\usepackage{xspace}
\usepackage[letterpaper,top=1in,bottom=1in,hmargin={1in,1in}]{geometry}
\numberwithin{equation}{section}
\DeclareRobustCommand\em
{\@nomath\em \ifdim \fontdimen\@ne\font >\z@
  \upshape \else \slshape \fi}

\usepackage{babel}
\makeatother
\begin{document}
%\nolinenumbers

\title{A Cognitive Model of an Epistemic Community: Mapping the Dynamics of Shallow Lake Ecosystems}

\author{Can Ozan TAN$^{1*}$ and Uygar \"{O}ZESMI$^{2}$}

\date{}

\maketitle
\noindent $^{1}$Canakkale University, Dept. of Biology, Terzioglu
Kampusu, 17020, Canakkale, TURKEY

\noindent $^{2}$Erciyes University, Dept. of Environmental Engineering,
38039 Kayseri, TURKEY

\noindent $^{*}$Corresponding Author: Presently at Boston University
Dept. of Cognitive and Neural Systems, 677 Beacon Street \#201, Boston,
MA 02215, USA. e-mail: tanc@cns.bu.edu Tel: +1-617-353-6741 Fax: +1-617-353-7755

\vspace{0.5in}


\vspace{0.5in}

\pagebreak

\begin{abstract}
We used fuzzy cognitive mapping (FCM) to develop a generic shallow
lake ecosystem model by augmenting the individual cognitive maps drawn
by 8 scientists working in the area of shallow lake ecology. We calculated
graph theoretical indices of the individual cognitive maps and the
collective cognitive map produced by augmentation. There were a total
of 32 variables with 113 connections in the collective cognitive map.
The graph theoretical indices revealed internal cycles showing non-linear
dynamics in the shallow lake ecosystem. The ecological processes were
organized democratically without a top-down hierarchical structure.
The most central variable in the collective map was submerged plants.
The strongest connections were suspended solid concentration decreasing
water clarity, phosphorus concentration increasing the phytoplankton
biomass, higher water clarity increasing submerged plants, benthivorous
fish biomass reducing submerged plants and increasing suspended solid
concentration, and submerged plants decreasing suspended solids.

The steady state condition of the generic model was a characteristic
turbid shallow lake ecosystem. The generic shallow lake ecosystem
model had the tendency to go into a turbid state since there were
no dynamic environmental changes that could cause shifts between a
turbid and a clearwater state, and the generic model indicated that
only a dynamic disturbance regime could maintain the clearwater state.
The model developed herein captured the empirical behavior of shallow
lakes, and contained the basic model of the Alternative Stable States
Theory. In addition, our model expanded the basic model by quantifying
the relative effects of connections and by extending it with 22 more
variables and 99 more weighted causal connections. On our expanded
model we ran 4 simulations: harvesting submerged plants, nutrient
reduction, fish removal without nutrient reduction, and biomanipulation.
Only biomanipulation, which included fish removal and nutrient reduction,
had the potential to shift the turbid state into clearwater state.
The structure and relationships in the generic model as well as the
outcomes of the management simulations were supported by actual field
studies in shallow lake ecosystems. Thus, fuzzy cognitive mapping
methodology enabled us to understand the complex structure of shallow
lake ecosystems as a whole and obtain a valid generic model based
on tacit knowledge of experts in the field. 
\end{abstract}

\clearpage
%\twocolumn

\section{Introduction}

Shallow lake ecosystems are one of the most numerous and one of the most vulnerable ecosystems in the world. Most shallow lakes are suffering from structural and functional changes and food web breakdown known as eutrophication as a result of excessive nutrient inputs from increased agricultural and urban activities, especially in highly populated areas (Moss, 1998). Until the early 1990's, the management and restoration strategies for eutrophic shallow lakes focused on decreasing inflowing and in-lake nutrients, mainly phosphorus concentrations (Carpenter, Kitchell \& Hodgson, 1985; Edmonson, 1985). However, eutrophic shallow lakes showed resilience and resistance to restoration (Sas, 1989; Scheffer et al, 1993a). The Alternative Stable States Theory showed that shallow lakes may be in two alternative stable states, a phytoplankton dominated turbid water state and a submerged plant dominated clearwater state (Scheffer et al. 1993a; Moss, 1990, 1998; Scheffer, 1998; Jeppesen, 1998). Each state is stable, in the sense that a shift from one state to the other requires a strong intervention from outside of the system. Thus, developing management plans for restoration and predicting the outcomes of those plans using mathematical modelling tools and computer simulations has been one of the key concerns of shallow lake ecologists and managers. 
So far, several models of shallow lake ecosystems have been developed. Many studies in the 1970s tried to model the eutrophication process in lakes using differential/difference equation models where parameters were empirically obtained (Ross, 1976, Lassiter and Kearns, 1977). These models assumed that the response of the system to the disturbance was in a linear fashion (Vollenweider, 1975), and were case-specific models. Later in the 1990's, more general models of shallow lakes were built, both aiming to explain the whole ecosystem structure (Scheffer et al., 1993a; Scheffer, 1998), and focusing on a particular compartment such as submerged plants (Scheffer, Bakema \& Wortelboer, 1993b). The model developed by Scheffer et al. (1993a) became the basic model, which could be modified for specific cases. Recently, Zhang et al. (2003) employed a dynamic structural model for a shallow lake ecosystem. Although this latter model was successful in predicting the dynamics of the system, it was also case-specific.

In this study, we use cognitive mapping to develop a shallow lake ecosystem model that is derived from the collective opinions and experiences of scientists who have been working on shallow lakes. The hypothesis is that this model will be a successful and detailed generic model of shallow lake ecosystems. We tested the results of this model with empirical results published in scientific literature. 

Cognitive maps are defined as qualitative models of how a given system operates. A cognitive map is based on defined variables, which can be physical quantities, or complex and abstract ideas, and the causal connections among them (\"{O}zesmi and \"{O}zesmi, 2004). Cognitive mapping is a useful tool to model complex systems such as shallow lake ecosystems. In addition, cognitive mapping allows for combining different expert opinions which result from different experiences in different physiogeographic contexts.

\section{Methods}

We collected the cognitive maps of 8 scientists and coded them as square matrices. To analyze the structure of the cognitive maps, we calculated the graph theoretical indices of the matrices. Then we augmented the matrices to obtain a collective cognitive map. The collective map was further used for simulations to test the outcomes of possible management options and the results were compared to the results of other studies in the literature to test the performance of the model. 

\subsection{Why use cognitive mapping?}

Cognitive mapping is a methodology where expert perception of complex systems can easily be obtained, combined and simulated. Cognitive mapping can model nonlinear relationships and feedbacks that are characteristics of many ecosystems. Although ecological applications are rare, the method has solid foundations. Cognitive maps are based on graph theory and digraphs (see Bang-Jensen and Gutin, 2002 for full treatment of the topic). Harary, Norman \& Cartwright (1965) showed how directed graphs can be used to study the structural properties of the empirical world. In our study, we considered shallow lake ecosystems as empirical worlds. Since the 1970s, numerous studies have used cognitive mapping for representing complex systems and for decision-making (Axelrod, 1976; Bauer, 1975; Bougon et al., 1977, Brown, 1992; Carley and Palmquist, 1992; Cossette and Audet, 1992; Hart, 1977; Klein and Cooper, 1982; Malone, 1975; Montazemi and Conrath, 1986; Nakamura, Iwai \& Sawaragi, 1982; Roberts, 1973). In the 1980s Kosko modified binary cognitive maps by applying fuzzy causal connections with real numbers in the range $[-1,1]$ to the connections, using the term of fuzzy cognitive map (FCM), and provided means to simulate these FCMs (Kosko, 1986, 1987, 1988, 1992a, 1992b).  The approach of FCM has been used in a wide variety of research, ranging from political developments (Taber, 1991) to data mining of the World Wide Web (Hong and Han, 2002; Lee et al., 2002).  \"{O}zesmi (1999) used cognitive maps to represent local knowledge systems as told by informants in different stakeholder groups, the main assumption being that individuals have cognitive models that are internal representations of a partially observed world (Bauer, 1975).  

The use of FCM in ecology has been rather limited. Known examples include Radomski and Goeman (1996) who analyzed possible management options for sports fisheries; Hobbs et al. (2002), who used FCM to define management objectives for Lake Erie ecosystem; \"{O}zesmi and \"{O}zesmi (2003), who analyzed the perceptions of different stakeholder groups about a lake ecosystem in order to develop a management plan; and Skov and Svenning (2003), who combined FCM with a GIS to use expert knowledge to predict plant habitat suitability in a forest.

\subsection{Data Collection}

In this study, we used a multi-step fuzzy cognitive mapping approach (\"{O}zesmi and \"{O}zesmi, 2004; a detailed description of the standard method used can be found in this methodological paper) to model the system dynamics of shallow lakes based on models drawn/mapped by 8 expert scientists in the discipline. The scientists were all male, with an average age of $37.9 \pm 7.3$ years. The question they were asked was 'What are the variables, factors determining the state of shallow lake ecosystems and how do they interact with each other?' They defined variables and drew causal relationships among these variables. In addition, they indicated the relative strength/weight and direction of the relationships using a scale varying between -1 and +1 with 0.25 increments. The sample size (i.e., the number of maps) was considered sufficient once the number of new variables leveled off as the maps were added together (\"{O}zesmi and \"{O}zesmi 2004, p. 48). The number of new variables added per new map leveled off at about 1 by the 8$^\mathrm{th}$ interview after 1000 Monte-Carlo simulations (figure 1) using EstimateS 6.0b1 (Colwell, 1997). Accumulation of variables indicated that 8 maps were sufficient to characterize the system under study.

\subsection{Data Analysis}

Once the maps were drawn, they were coded as square (adjacency) matrices according to graph theory in the form $A(D)=[a_{ij}]$ (Harary et al., 1965).  The variables $v_i$ are listed on the vertical axis and $v_j$  on the horizontal axis. When a connection exists between two variables the value is coded in the square matrix (between -1 and 1). 

Then the structure of the maps was analyzed using indices derived from graph theory. We calculated the indegrees and outdegrees of each variable for each individual map. Indegree is defined as the column sum of absolute values of a variable and shows the cumulative strength of variables entering the variable. Outdegree is defined as the row sum of absolute values of a variable and shows the cumulative strengths of connections exiting the variable.  In addition, we determined the variables with positive outdegree and zero indegree, with positive indegree and zero outdegree and with non-zero indegree and outdegree, which represent transmitter, receiver and ordinary variables, respectively. For each individual map to determine its network structure, we also calculated the centrality, which is the sum of indegree and outdegree and indicate the contribution of that variable to the cognitive map; complexity index, which is the ratio of number of receiver variables to that of the transmitter variables; connectivity as the number of connections; density, which is calculated as a function of connectivity, and shows how connected or sparse the maps are; distance, which is the average causal link between variables and hierarchy index, which is a function of the outdegrees and number of variables in a given map and shows how democratic the map is, (see \"{O}zesmi and \"{O}zesmi, 2004, p. 50-51. for formulas).

Individual maps were later augmented (matrices were added together in an overlapping manner (Kosko, 1987, 1992a,b)) to obtain a collective cognitive map (social map) on the dynamics of shallow lake ecosystems. First an augmented matrix is created that includes all the variables from all the individual cognitive maps.  Second each individual cognitive map is coded into an augmented matrix and the individual cognitive maps are added together using matrix addition to create a social cognitive map.  As the maps are added together, conflicting connections with opposite signs will decrease the causal relationship, while agreement reinforces causal relationships, forming a collective map, which is considered a consensus opinion of the scientists.  While augmenting, equal weight was given to the map of each scientist.  

The collective cognitive map was used in simulating system behavior and to run management simulations. Simulations were made by multiplying a vector of initial states of variables ($\mathrm{I^n}$) with the adjacency (square) matrix A of the collective map.  Usually a threshold function or a transformation by a bounded monotonic increasing function is applied to the result of the matrix multiplication, $\mathrm{I^n \times A}$, at each simulation time step (Kosko 1987; Kosko 1992b, \"{O}zesmi and \"{O}zesmi 2004, p. 55).  We used a logistic function $(1+e^{-x})^{-1}$ to transform the results into the interval $[0,1]$. The resulting transformed vector was then repeatedly multiplied by the adjacency matrix and transformed until the system converged to a fixed point. We produced 10 random initial state vectors and ran the model to determine the stability of the collective map without any management. Later management options were simulated by clamping given variables through the multiplication process.  Relevant variables were selected based on common practice shallow lake management interventions and these were simulated. The simulation results were compared with results of empirical studies.  All computations were carried out using subroutines implemented in Octave programming language version 2.0.16 on a Linux OS.

\section{Results} 

The mean number of variables in the individual cognitive maps of the shallow lake ecosystems drawn by 8 scientists was $14.6 \pm 3.3$ SD, while there were $20.9 \pm 5.7$ connections on average between the variables that they defined. There were a total of 32 variables with 113 connections in the collective cognitive map obtained by augmenting the 8 individual maps (table 1). 

Mean values of graph theoretical indices of the individual maps and the collective map are shown in table 1. Scientists have $3.8 \pm 2.5$ SD transmitter variables in their individual maps, whereas in their collective cognitive map, there are a total of 6 transmitter variables. These were external nutrient load, micropollutants, dissolved organic carbon concentration, temperature, fetch and shoreline slope. There were about the same number of receiver variables  (utility variables), $3.4 \pm 2.1$ SD in the individual maps, whereas only one in the collective map (sediment phosphorus release). There were $10.6 \pm 3.1$ ordinary variables (means) in the individual maps, and 25 in the collective map. 

Looking at the cognitive maps, we observe that there are on average $1.56 \pm 0.44$ SD causal chains in the individual maps, whereas there are 2.79 causal connections in the collective map. The complexity, density and hierarchy indices of the cognitive maps are low (table 1). 

The most frequently mentioned variables that are recurrent in the 8 cognitive maps are given in table 2. Suspended solids were mentioned in each map, and water clarity, zooplanktivorous fish, submerged plants, benthivorous fish and phosphorus concentration were mentioned in 7 of the 8 maps. 

The most central variable in the collective map was submerged plants (figure 2). Moreover, submerged plants have as much effect on other variables (outdegree: 2.8) as it is affected by others (indegree: 3.0). 

In figure 2, variables are ordered according to their centralities. Variables shown in the figure have centrality higher than 1.0, and were mentioned by more than half of the participating scientists. Suspended solid concentrations, phytoplankton biomass, water clarity and zooplankton biomass were more affected by other variables than affecting others (indegree $>$ outdegree), while the situation was the opposite for phosphorus concentration, benthivorous and planktivorous fish biomass, waves and wind action, benthic biomass and water level. 

We drew the cognitive interpretation diagram (CID, figure 3) according to \"{O}zesmi and \"{O}zesmi (2004, p. 53). This diagram transforms the complex "spaghetti and meat balls" graph of the collective map into an understandable format where connections with weights more than 0.25 are drawn, scaled to their actual weights. Onto our CID, we superimposed the basic model of Scheffer et al. (1993a). Variables mentioned also by Scheffer et al., are drawn in two concentric circles, compared to those with only one circle, which are our additional variables. Binary connections that are also present in Scheffer et al. (1993a) have a circle at the arrowheads and have all the same signs as ours. Connections missing in our model (from submerged plants to allelophatic substances, and from there to phytoplankton, from submerged plants to zooplankton-fish interaction) have a thin line with a circle at the pointing end. 

Accordint to the CID (figure 3), the strongest connections are suspended solid concentration decreasing water clarity, phosphorus concentration increasing the phytoplankton biomass and higher water clarity increasing submerged plants (see table 3 for actual weights). These connections are followed in strength by benthivorous fish biomass reducing submerged plants and increasing suspended solid concentration, submerged plants decreasing suspended solids and zooplankton density reducing the phytoplankton biomass (table 3). 

We determined the steady state condition of the system before any management option was considered, in order to be able to see the tendency of the ecosystem based on the collective cognitive map. We ran the generic shallow lakes ecosystem model with 10 different random initial states for all variables between 0 and 1 from a uniform distribution. We also ran an initial state vector where all variables were set to 1, as well as with a real data set compiled from Lake Eymir database (Burnak and Beklioglu, 2000; Tan, 2002). In all of these no management simulations, the system reached steady state after 10 - 12 iterations, where the states were the same to $10^{-6}$ digits. A one-way ANOVA test did not show any significant differences ($\mathrm{F}_{11,341}: 0.000, p: 1.000$). The mentioned steady state condition of the system is given in figure 4. The steady state condition of the collective generic model is a characteristic turbid shallow lake ecosystem where primary production was governed by phytoplankton and benthic primary production. As figure 4 clearly shows, phytoplankton biomass, phosphorus, and suspended solids concentrations are the highest variables, and therefore not surprisingly, water transparency and submerged plants are almost 0. 

Following the determination of system's steady state, management scenarios are developed based on common interventions in the field of shallow lake restoration. These scenarios were: A) Harvesting submerged plants; B) Nutrient reduction; C) Fish removal without nutrient reduction; and D) Biomanipulation involving both fish removal and nutrient reduction. The results of the managemnt simulations are shown in figure 5. We looked how much the most central variables (water clarity, phosphorus concentration, phytoplankton biomass, zooplankton density, planktivorous fish biomass, benthivorous fish biomass, benthic biomass, submerged plants, suspended solid concentration, water, level, and waves/wind action) changed from their steady state values. A negative value indicates a reduction in the variable state compared to its steady state, while a positive value reflects an increase in the variable state. 

\textit{Harvesting Submerged Plants.} Submerged plants are usually harvested in recreational lakes where it is considered a nuisance to boating and angling as well as aesthetically displeasing. When submerged plants are harvested (figure 5-A), the main change observed is an increase in the suspended solid concentration. While the phosphorus concentration and phytoplankton biomass slightly increased, zooplankton density decreased. An interesting result is the increase in the wave and wind action, revealing the importance of submerged plants for stabilizing the wave action, and thus, decreasing the suspended solids. When we look at the scale of change in variables under this scenario, compared to other management options (figure 5-B, C, D), we see that it is relatively low. The low relative change is reasonable, considering that the system at its steady state is already in turbid water conditions with very low water transparency and few submerged plants (figure 4). 

\textit{Nutrient Reduction.} In this simulation (figure 5-B), nutrient reduction leads to a strong decrease in the phytoplankton biomass, a reduction in the concentration of suspended solids, and a decrease in the planktivorous and benthivorous fish biomass.

\textit{Fish Removal without Nutrient Reduction.} When the zooplanktivorous fish were removed, naturally the zooplankton density increased and the phytoplankton biomass decreased. By also removing benthivorous fish, the suspended solids decreased considerably in our model (figure 5-C). 

\textit{Bimanipulation.} In the last simulation (figure 5-D) benthivorous and planktivorous fish were removed and nutrients were also reduced. In comparison to all the other scenarios we observed a much higher change in the variables and a shift towards a clearwater state. For the first time submerged plants increased considerably as well as water clarity and zooplankton density, while phytoplankton biomass decreased. Suspended solids also decreased mainly as an outcome of the decreased benthivorous fish stock, as suggested in scenario 3 (fish removal). 

\section{Discussion}

\subsection{Graph Theoretical Indices}

The low mean number of transmitter variables in the cognitive maps of scientists indicates that they define relatively few outside controls (forcing functions). The mean number of receiver variables, being also low, and the high number of ordinary variables in the cognitve maps further show that the shallow lake ecosystem is predominantly governed by circular internal ecological processes. These internal processes define how the ecosystem will behave. Furthermore, the low values of the complexity, density and hierarchy indices, considered among with the mean number of causal connections, indicates that according to the scientists, the processes in the shallow lake ecosystem are organized "democratically" without showing a top-down hierarchical structure, and there are many internal cycles showing non-linear dynamics.

\subsection{Variable characteristics}

Although the most mentioned variables show what is in the minds of the participating scientists, we can understand the most important variables that control the dynamics of the shallow lake ecosystem by looking at the centrality of variables in the collective cognitive map. The indegree and outdegree, which add up to centrality, will indicate how much the variable is affected by other variables or how much it affects the others. Thus, the differences in indegrees and outdegrees of the variables are important for inferring the relative role of that variable in the ecosystem.

The most central variable in the collective map is found to be the submerged plants, assigning a key central role to submerged plants. This result corroborates the field studies and alternative stable states theory, where submerged plants are the key components in determining the ecosystem dynamics in a shallow lake (Scheffer et al., 1993a; Jeppesen, 1998). Moreover, close values of indegree and outdegree of submerged plants indicated that they have as much effect on other variables as they are affected by others. This result is similar to the figure of the basic shallow lake ecosystem model by Scheffer et al.(1993a), where vegetation has a central position and both affects and is affected by other variables. 

Variable centralities in the cognitive map (figure 2) revealed two dominant groups of variables. Accordingly, suspended solid concentrations, phytoplankton biomass, water clarity and zooplankton biomass were more affected by other variabes than affecting others, whereas the oppoite is true for phosphorus concentration, benthivoros and planktivorous fish biomass, waves and wind action, benthic biomass and water level. The latter group will act more like forcing functions, influencing ecosystem processes (high outdegrees) while being less influenced by other ecosystem components (low indegrees). While these results are not surprising according to our current knowledge of shallow lake ecosystems, it is surprising that the generic model captures the relationship of the benthi-planktivorous fish biomass with other components of the ecosystem. As the model shows, fish will be less affected by other variables having a much longer life span compared to other organisms, such as zooplankton. During their longer life span, fish will have a lasting effect on the shallow lake ecosystem dynamics. 

\subsection{Cognitive Interpretation Diagram of the Shallow Lake Ecosystem}

Cognitive interpretation diagram (CID, figure 3) revealed the fact that basic shallow lake model introduced by Scheffer et al. (1993a) is consistent with and included in our model. In our model, allelophatic substances do not have a strong effect in the system. The advantage of our model is that all connections regardless whether they are included in Scheffer et al. (1993a) or not, have weights associated with them showing their relative importance in the ecosystem dynamics. 

The strength and pattern of the connections in our model, according to CID, are also consistent with actual field studies in shallow lake ecosystems and theoretical conclusions derived from these (Jeppesen, 1998; Moss, 1998; Scheffer, 1998). Our model contains the basic model of Scheffer et al.(1993a).  This is not surprising because all of the scientists that contributed their cognitive maps certainly knew of Scheffer's model and were influenced by it.  However the collective cognitive map improves Scheffer's basic model by quantifying the relative effects of connections and adding 22 more variables and 99 causal connections.   

\subsection{Simulations Using Collective Fuzzy Cognitive Map}

The multi-step fuzzy cognitive mapping approach (\"{O}zesmi and \"{O}zesmi, 2004) enables us to run the collective cognitive map as a generic shallow lake ecosystem simulation model. We can model the system as it is without any management, run simulations of management options by fixing state values of relevant variables and find the impact of those management options by looking at deviations from the steady state. One should be careful in interpreting results obtained from simulations of fuzzy cognitive maps, because these are not system dynamics models in the classical sense, yielding directly measurable results. Rather, the results are qualitative indicators showing whether a variable increases or decreases relative to its steady state or to other variables in the system. 

\subsubsection{Steady State Condition}

Although we did not indicate anything regarding the state of the system to the 8 scientists who drew their maps, their collective generic model, where there is no human management, ends up at a steady state condition that is in turbid water state. Moreover, we did a simulation to test this steady state, excluding all the possible variables that may be considered as outside disturbances to the system (external nutrient load, sediment phosphorus release, micropollutants, bioturbation, and waves and wind action), and the system still stabilized on a turbid water state. This might be because these scientists are experts on management and restoration of shallow lake ecosystems which are usually already in turbid water state. Alternatively, this result raises the question whether the natural tendency of most shallow lake ecosystems is to stay in a turbid state if conditions are stable over a long period of time. Maybe, what maintains or causes shifts from turbid to clearwater state is a dynamic environment in which variables change based on conditions outside the system. In other words, is it a dynamic disturbance regime that maintains the clearwater state in shallow lake ecosystems? We know that Fox (1979), Connell (1979), and Reynolds, Padisak \& Sommer (1993) postulate in the intermediate disturbance hypothesis that high species diversity in an assemblage is maintained by fluctuations in the environmental conditions occurring within the generation time of those organisms forming the assemble. Similarly, our model suggests that diversity at the ecosystem level is supported by outside disturbances. This, in turn, would contribute to diversity at the species level. 

Our shallow lake ecosystem model represents the ecosystem processes as they are. Thus applying different management options can be considered as disturbing the system in a directed manner, taking advantage of the system characteristics to maintain a clearwater state. Maintaining a clearwater state is preferred for reasons of productivity, biodiversity and aesthetic concerns. But if all shallow lakes in the world were maintained in a clearwater state this would be a loss in total biodiversity. However, today, human economic activities cause most shallow lakes to turn into turbid ecosystems. Therefore, we also explored different management options that can provide for a clearwater state.

\subsubsection{Simulation of Management Scenarios for Shallow Lakes}

The collective cognitive map of the shallow lake ecosystem can be considered an a priori model, because no empirical studies were used to construct or parameterize the model. Of course, one could argue that the scientists who contributed to the generic model have done so based on their own empirical work and knowledge derived from the existing literature. However, the models that they shared with us are based on tacit knowledge that does not involve explicitly measurable relationships but qualitative judgments on cause and effect relationships between variables. By comparing the performance of this a priori qualitative model directly with empirical results, we can continue to test its generic validity. 

\textit{Harvesting Submerged Plants.} This scenario agrees with the general conclusions of field studies where the loss of submerged plants was followed by an increase in suspended solid concentrations mainly because of an increase in wind and wave action (Bostr\"{o}m, Jansson \& Forsberg, 1982). Moreover, Breukelaar et al.(1994) reported that the loss of submerged plants leads to an increase in the benthic fish biomass. They concluded that it is easier for benthivorous fish to feed when there are no submerged plants. This is exactly what we have observed in our simulation (figure 5-A). Moreover, it is known that the submerged plants enhance the zooplankton density by providing refuge to zooplankton against zooplanktivorous fish predation (Timms and Moss, 1984). As submerged plants have been removed in our simulation, we observe a drop in zooplankton density. The expected increase in phytoplankton biomass due to a reduction in grazing zooplankton density is also observed in this simulation. 

\textit{Nutrient Reduction.} The strong decrease in the phytoplankton biomass and reduction in the concentration of suspended solids is probably as a result of the observed decrease in the planktivorous and benthivorous fish biomass. In agreement with alternative stable states theory (Scheffer et al., 1993a), despite the strong decrease in phytoplankton biomass, we do not observe any improvement in the water clarity and submerged plants which can be attributed to the resilience of the system. Field studies have corroborated this resilience repeatedly. (Scheffer, 1998; Breukelaar et al., 1994; Jeppesen et al., 1997, 1999; Jeppesen, 1998; Beklioglu, Carvalho\& Moss, 1999). 

\textit{Fish Removal without Nutrient Reduction.} Our model is in agreement with Scheffer (1998). When th zooplanktivorous firs are removed, naturally the zooplankton density increases, and as a consequence, the phytoplankton biomass decreases due to intensified grazing (Scheffer, 1998). Also in our model, suspended solids were decreased by removing benthivorous fish species. Benthivorous fish species are known to enhance suspended solids concentration due to their feeding action on the sediment (Meijer et al., 1990; Breukelaar et al., 1994; Sidorkewicj, Lopez Cazorla \& Fernandez, 1996). When benthivorous fish feed on the sediment, they stir up the phosphorus bound in the sediment into the water column. (Horpilla and Kairesalo, 1990; Breukelaar et al., 1994). We observed this effect by a slight decrease in the concentration of phosphorus when benthivorous fish were removed. 

We did not observe,however, an increase in water clarity despite the reduction in suspended solids and phytoplankton biomass when only fish were removed. The same has been observed in empirical studies (Jeppesen et al, 1997, 1999; Hansson et al., 1998; Beklioglu et al., 1999). The alternative stable states theory (Scheffer et al., 1993a) explains this phenomenon as the resilience of the system to shift into a clearwater state. The resilience of the turbid water state is a result of the delayed development of submerged plants. Before submerged plants can get truly established in the lake and their stabilizing buffer mechanisms can take effect, high nutrient concentrations in the water favor phytoplankton redevelopment (James \& Barko, 1990; Scheffer et al., 1993a; Weisner, Strand \& Sandsten, 1997; van Donk \& van de Bund, 2002).

\textit{Biomanipulation.} The shift to clearwater state in this scenario is consistent with the biomanipulation literature (Shapiro, Lamarra \& Lynch, 1975; Benndorf et al., 1988; Hosper, 1989; Jeppesen et al., 1990; Carpenter and Kitchell, 1993; Philips and Moss, 1994; Hansson et al., 1998; Meijer, 2000). Moreover, the action of waves and wind decreased, most probably as an outcome of submerged plant development (Bostr\"{o}m et al., 1982).

In summary, we used cognitive mapping to develop a generic shallow lake ecosystem model that is derived from the collective tacit knowledge and experiences of 8 scientists who have been working on shallow lakes. There were a total of 32 variables with 113 connections in the collective cognitive map. The graph theoretical indices show that the shallow lake ecosystem is predominantly governed by internal cycles showing non-linear dynamics. The ecological processes are organized democratically without showing a top-down hierarchical structure.

Submerged plants were the most central variable in the collective map. The strongest connections in decreasing order were: suspended solid concentration decreasing water clarity; phosphorus concentration increasing the phytoplankton biomass; water clarity improving submerged plants; benthivorous fish biomass reducing submerged plants and increasing suspended solid concentration; submerged plants decreasing suspended solids; and zooplankton density reducing the phytoplankton biomass. The steady state condition of the generic model was a characteristic turbid shallow lake ecosystem where primary production was governed by phytoplankton and benthic primary producers. The turbid state might be because participating scientists are experts on management and restoration of shallow lake ecosystems that generally show a high trophic state. Alternatively, most shallow lake ecosystems have the natural tendency to stay in a turbid state if conditions are stable over a long period of time. Shifts between a turbid and a clearwater state may be the result of a dynamic environment in which variables change in relation to the conditions outside the system. Our model indicates that only a dynamic disturbance regime could maintain the clearwater state in shallow lake ecosystems similarly to how species diversity is maintained in the intermediate disturbance hypothesis. Diversity at the ecosystem level created by outside disturbances can contribute to diversity at the species level in large physiogeographic scales. 

The generic FCM model captured all the behavior predicted by the Alternative Stable States Theory developed for shallow lakes ecosystems. Our model contains the basic model of Scheffer et al. (1993a) but expands it by quantifying the relative effects of connections and adds 22 more variables and 99 more weighted causal connections. On our expanded model we ran 4 simulations: harvesting submerged plants, nutrient reduction, fish removal without nutrient reduction, and biomanipulation. Harvesting submerged macrophytes did not show a strong impact on the system because the steady state of the system was already turbid. This simulation showed that submerged plants have an important effect even in a turbid state where they are minimal. Similarly, neither nutrient reduction nor fish removal was enough by itself for the lake to shift toward a clearwater state. However, fish removal accompanied with nutrient reduction (biomanipulation scenario) showed that this strategy is applicable for restoring turbid water systems. Not only the structure and relations in the generic model were supported by actual field studies in shallow lake ecosystems and theoretical conclusions derived from these (Jeppesen, 1998; Moss, 1998; Scheffer, 1998) but also all the outcomes of the 4 management simulations were supported by other empirical results in literature.

Cognitive mapping methodology also enabled us to explore a complex system in detail and gain insights about how the ecosystem functions without being trapped into the specificity of sites but be able to generalize from them.  The model produced outcomes consistent with empirical results. The behavior of the model suggested that shallow lake ecosystems have a tendency to become turbid unless there are dynamic forcing functions changing central variables. This theoretical insight is consistent with the theory that there is a tendency for an ecosystem to deteriorate unless the dynamics of outside forces rejuvenate critical central variables (Connell, 1979; Fox, 1979; Reynolds et al., 1993).

%\pagebreak

\section*{Acknowledgements}

We thank -Paul Boers, Laurence Carvalho, Hugo Coops, Luigi Naselli-Flores, Marten S\o ndergaard, John Strand, Can O. Tan, and Egbert van Nes who kindly provided their cognitive maps, which made this study possible. We also thank Luigi Naselli-Flores and Egbert van Nes for reviewing drafts of this manuscript.

%\pagebreak

\section*{References}

\noindent
Axelrod, R. (1975) \textit{Structure of Decision: The Cognitive Maps of Political Elites.} Princeton University Press, Princeton, NJ, USA.

\noindent
Bang-Jensen, J. and Gutin, G. (2002) \textit{Digraphs: Theory, Algorithms and Applications.} Springer Verlag, London.

\noindent
Bauer, V. (1975) Simulation, evaluation and conflict analysis in urban planning. In \textit{Portraits of Complexity, Applications of Systems Methodologies to Societal Problems} (Ed. M.M. Baldwin) pp.179-192. Battelle Institute, Columbus, Ohio. 

\noindent
Beklioglu M., Carvalho L. \& Moss B. (1999) Rapid recovery of a shallow hypertrophic lake following sewage effluent diversion: lack of a chemical resilience. \textit{Hydrobiologia}, \textbf{412}, 5-15.

\noindent
Benndorf, J., Kneschke, H., Kossatz K. \& Penz, E. (1988) Manipulation of the pelagic food web by stocking with predacious fishes. \textit{International Review Gesamten Hydrobiologie}, \textbf{69}, 407-428. 

\noindent
Bostr\"{o}m, B., Jansson, M. \& Forsberg, C. (1982) Phosphorus release from lake sediments. \textit{Archive fuer Hydrobiologie/Bezihungs Ergebnisse der Limnologie}, \textbf{18}, 5-59.

\noindent
Bougon, M., Weick, K. \& Binkhorst, D. (1977) Cognition in organizations: an analysis of the Utrecht Jazz Orchestra. \textit{Administrative Science Quarterly}, \textbf{22}, 606-639.

\noindent
Breukelaar, W.A., Lammens, H.R.R.E., Breteler, J.G.P. \& Tatrai, I. (1994) Effects of benthivorous bream (\textit{Abramis brama}) and carp (\textit{Cyprinus carpio}) on sediment resuspension and concentrations of nutrients and chlorophyll-a. \textit{Freshwater Biology}, \textbf{32}, 113-121.

\noindent
Brown, S.M. (1992). Cognitive mapping and repertory grids for qualitative survey research: Some comparative observations. \textit{Journal of Management Studies}, \textbf{29}, 287-307.

\noindent
Burnak, S.L. \& Beklioglu, M. (2000) Macrophyte dominated clearwater state of Lake Mogan. \textit{Turkish Journal of Zoology}, \textbf{24}, 305-313.

\noindent
Carley, K. \& Palmquist, M. (1992) Extracting, representing, and analyzing mental models. \textit{Social Forces}, \textbf{70}, 601-636.

\noindent
Carpenter, R.S., Kitchell, J.F. \& Hodgson, R.J. (1985) Cascading trophic interactions and lake productivity. \textit{Bioscience}, \textbf{35}, 634-638.

\noindent
Carpenter, S.R. \& Kitchell J.F. (1993) \textit{The trophic cascade in lakes}. Cambridge University Press, Cambridge.

\noindent
Colwell, R.K. (1997) \textit{Estimate S: Statistical estimation of species richness and shared species from samples} Version 5. User's Guide and application published at: http://viceroy.eeb.u-conn.edu/estimates.

\noindent
Connell, J.H. (1979) Intermediate-disturbance hypothesis. \textit{Science}, \textbf{204}, 1345-1345.

\noindent
Cosette, P. \& Audet, M. (1992) Mapping of an idiosyncratic schema. \textit{Journal of Management Studies}, \textbf{29}, 325-347.

\noindent
Edmonson, W.T. (1985) Recovery pf Lake Washington from eutrophication. Pages 228-234 in \textit{Proceedings from lakes pollution and recovery}, Rome, 15-18 April.

\noindent
Fox, J.F. (1979) Intermediate-disturbance hypothesis. \textit{Science}, \textbf{204}, 1344-1345.

\noindent
Hansson, L.A., Annadotter, H., Bergman, E., Hamrin, S.F., Jeppesen, E., Kairesalo, T., Luokkanen, E., Nilson, P.A., S\o ndergaard, M. \& Strand, J. (1998) Biomanipulation as an application of food chain theory: constraints, synthesis and recommendations for temperate lakes. \textit{Ecosystems}, \textbf{1}, 558-574.

\noindent
Harary, F., Norman, R.Z. \& Cartwright, D. (1965) \textit{Structural Models: An Introduction to the Theory of Directed Graphs}, John Wiley \& Sons, New York.

\noindent
Hart, J.A. (1977) Cognitive maps of three Latin American policy makers. \textit{World Politics}, \textbf{30}, 115-140.

\noindent
Hobbs, B.F., Ludsin, S.A., Knight, R.L., Ryan, P.A., Biberhofer, J. \& Ciborowski, J.J.H. (2002) Fuzzy cognitive mapping as a tool to define management objectives for complex ecosystems. \textit{Ecological Applications}, \textbf{12} 1548-1565.

\noindent
Hong, T. \& Han, I. (2002) Knowledge-based data mining of news information on the internet using cognitive maps and neural networks. \textit{Expert Systems with Applications}, \textbf{23}, 1-8.

\noindent
Horpilla, J. \& Kairesalo T., (1990) A fading recovery: the role of roach (\textit{Rutilus rutilus} L.) in maintaining high phytoplankton productivity and biomass in Lake Vesijarvi, Southern Finland. \textit{Hydrobiologia}, \textbf{200/201}, 154-163.

\noindent
Hosper, S.H. (1989) Biomanipulation, new perspectives for restoring shallow, eutrophic lakes in the Netherlands. \textit{Hydrobiological Bulletin}, \textbf{23}, 5-10.

\noindent
James, W.F. \& Barko, J.W. (1990) Macrophyte influences on the zonation of sediment accretion and composition in a north-temperate reservoir. \textit{Archive fuer Hydrobiologie}, \textbf{120}, 129-142.

\noindent
Jeppesen, E. (1998) \textit{The ecology of shallow lakes - Trophic interactions in the pelagial}. Doctor's dissertation (DSc). NERI Technical Report No. 247, Silkeborg.

\noindent
Jeppesen, E., Jensen, J.P., Kristensen, P., S\o ndergaard, M., Mortensen, E., Sortkjcer, E. \& Olrik, K. (1990) Fish manipulation as a lake restoration tool in shallow, eutrophic temperate lakes 2: threshold levels, long-term stability and conclusion. \textit{Hydrobiologia}, \textbf{200/201}, 219-227.

\noindent
Jeppesen, E., S\o ndergaard, M., Lauridsen, T.L., Pedersen, L.J. \& Jensen, L. (1997) Top-down control in freshwater lakes: the role of nutrient state, submerged macrophytes and water depth. \textit{Hydrobiologia}, \textbf{342/343}, 151-161.

\noindent
Jeppesen, E., Jensen, J.P., S\o ndergaard, M. \& Lauridsen, T.L. (1999) Trophic dynamics in turbid and clearwater lakes with special emphasis on the role of zooplankton for water clarity. \textit{Hydrobiologia}, \textbf{408/409}, 217-231.

\noindent
Klein, J.H. \& Cooper, D.F. (1982) Cognitive maps of decision-makers in a complex game. \textit{Journal of the Operational Research Society}, \textbf{33}, 63-71.

\noindent
Kosko, B. (1986) Fuzzy cognitive maps. \textit{International Journal of Man-Machine Studies}, \textbf{1}, 65-75.

\noindent
Kosko, B. (1987) Adaptive inference in fuzzy knowledge networks. Pages 261-268 in \textit{Proceedings of the First IEEE International Conference on Neural Networks (ICNN-86)}, San Diego, California.

\noindent
Kosko, B. (1988) Hidden patterns in combined and adaptive knowledge networks. \textit{Proceedings of the First IEEE International Conference on Neural Networks (ICNN-86)}, \textbf{2}, 377-393.

\noindent
Kosko, B. (1992a) Fuzzy associative memory systems, Pages 135-162 in A. Kandel (ed.) \textit{Fuzzy Expert Systems}, CRC Press, Boca Raton.

\noindent
Kosko,B. (1992b) \textit{Neural Networks and Fuzzy Systems: A Dynamical Systems Approach to Machine Intelligence.} Prentice-Hall, Englewood Cliffs, New Jersey.

\noindent
Lassiter, R.R. \& Kearns, D.K. (1977) Phytoplankton population changes and nutrient fluctuations in a simple aquatic ecosystem model. In \textit{Modeling the eutrophication process} (Eds. E.J. Middlebrooks, D.H. Falkenborg \& T.E. Maloney) $2^{\mathrm{nd}}$ edition, pp. 228. Ann Arbor Science Publishers. 

\noindent
Lee, K.C., Kim, J.S., Chung, N.H. \& Kwon, S.J. (2002) Fuzzy cognitive map approach to web-mining inference amplification. \textit{Expert Systems with Applications}, \textbf{22}, 197-211.

\noindent
Malone, D.W. (1975) An introduction to the application of interpretive structural modeling. In \textit{Portraits of Complexity: Applications of Systems Methodologies to Societal Problems} (Ed. M.M. Baldwin), pp. 119-126. Battelle Institute, Columbus, Ohio.

\noindent
Meijer, M. -L. (2000) \textit{Biomanipulation in the Netherlands: 15 years of experience}. Ministry of Transport, Public Works and Water Management, Institute for Inland Water Management and Waste Water Treatment (RIZA), Leystad, the Netherlands.

\noindent
Meijer, M. -L., de Haan, M.W., Breukelaar, A.W. \& Buiteveld, H. (1990) Is reduction of benthivorous fish an important cause of high transparency following biomanipulation in shallow lakes? \textit{Hydrobiologia}, \textbf{200/201}, 301-315.

\noindent
Montazemi, A.R. \& Conrath, D.W. (1986) The use of cognitive mapping for information requirements analysis. \textit{MIS Quarterly}, \textbf{10}, 45-55.

\noindent
Moss, B. (1990) Engineering and biological approaches to the restoration from eutrophication of shallow lakes in which aquatic plant communities are important components. \textit{Hydrobiologia}, \textbf{200/201}, 367-378.

\noindent
Moss, B. (1998) \textit{Ecology of Freshwaters: Man and Medium and Past to Future}. $\mathrm{3^{rd}}$ edition. Blackwell Science, Oxford.

\noindent
Nakamura, K., Iwai, S. \& Sawaragi,T. (1982) Decision support using causation knowledge base. \textit{IEEE Transactions on Systems, Man and Cybernetics SMC}, \textbf{12}, 765-777.

\noindent
\"{O}zesmi, U. (1999) Modeling ecosystems from local perspectives: Fuzzy cognitive maps of the Kizilirmak Delta Wetlands in Turkey. \textit{1999 World Conference on Natural Resource Modelling}, 23-25 June 1999, Halifax, Nova Scotia, Canada.

\noindent
\"{O}zesmi, U. \& \"{O}zesmi, S.L. (2003) A participatory approach to ecosystem conservation: Fuzzy cognitive maps and stakeholder analysis in Uluabat Lake, Turkey. \textit{Environmental Management}, \textbf{31}, 518-531. 

\noindent
\"{O}zesmi, U. \& \"{O}zesmi, S.L. (2004) Ecological Models based on People's Knowledge: A Multi-Step Fuzzy Cognitive Mapping Approach. \textit{Ecological Modeling}, \textbf{176}, 43-64.

\noindent
Phillips, G. \& Moss, B. (1994) \textit{Is biomanipulation a useful technique in lake management?} R \& D note 276, National Rivers Authority, UK.

\noindent
Radomski, P.J. \& Goeman, T.J. (1996) Decision making and modeling in freshwater sport-fisheries management. \textit{Fisheries}, \textbf{21}, 14-21.

\noindent
Reynolds C.S., Padisak, J. \& Sommer, S. (1993) Intermediate disturbance in the ecology of phytoplankton and the maintenance of species-diversity - a synthesis. \textit{Hydrobiologia},  \textbf{249}, 183-188

\noindent
Roberts, F.S. (1973) Building and analyzing an energy demand signed digraph. \textit{Environment and Planning}, \textbf{5}, 199-221.

\noindent
Ross, G.G. (1976) Plankton modelling in the Bay of Villefranche. \textit{Journal of Theoretical Biology}, \textbf{56}, 381-399.

\noindent
Sas, H. (1989) \textit{Lake restoration by reduction of nutrient loading: Expectations, experiences, extrapolations.} St. Augistin 1: Akademia Verl. Richarz. ISBN 3-88345379-X.

\noindent
Scheffer, M.  (1998) \textit{Ecology of shallow lakes.} Chapman \& Hall, London.

\noindent
Scheffer, M., Hosper, S.H., Meijer, M. -L., Moss, B. \& Jeppesen, E. (1993a) Alternative equilibria in shallow lakes. \textit{Trends in Ecology and Evolution}, \textbf{8}, 275-279.

\noindent
Scheffer, M.,  Bakema, A.H. \& Wortelboer, F.G. (1993b) MEGAPLANT: a simulation model of the dynamics of submerged plants. \textit{Aquatic Botany}, \textbf{45}, 341-356.

\noindent
Sidorkewicj, N.S., Lopez Cazorla, A.C. \& Fernandez, O.A. (1996) The interaction between \textit{Cyprinus carpio} L. and \textit{Potamogeton pectinatus} L. under aquarium conditions. \textit{Hydrobiologia}, \textbf{340}, 271-275.

\noindent
Shapiro, J., Lamarra, V. \& Lynch, M. (1975) Biomanipulation: an ecosystem approach to lake restoration. In \textit{Water Quality Management through Biological Control} (Eds. P.L. Brezonik \& J.L. Fox) pp 85-96. Rep. No. ENV-07-75-1, University of Florida, Gainesville.

\noindent
Skov, F. \& Svenning, J. -C. (2003) Predicting plant species richness in a managed forest. \textit{Forest Ecology and Management}, \textbf{6200}, 1-11.

\noindent
Taber, W.R. (1991) Knowledge processing with fuzzy cognitive maps. \textit{Expert Systems with Applications}, \textbf{2}, 83-87.

\noindent
Tan, C.O. (2002) \textit{The roles of hydrology and nutrients in alternative equilibria of two shallow lakes of Anatolia, Lake Eymir, and Lake Mogan: Using monitoring and modelling approaches}. M.Sc. Thesis. METU, Ankara, Turkey.

\noindent
Timms, R.M. \& Moss, B. (1984) Prevention of growth of potentially dense phytoplankton by zooplankton grazing, in the presence of zooplanktivorous fish, in a shallow wetland system. \textit{Limnology and Oceanography}, \textbf{29}, 472-486.

\noindent
van Donk, E. \& van de Bund, W.J. (2002) Impact of submerged macrophytes including charophytes on phyto- and zooplankton communities: allelopathy versus other mechanisms. \textit{Aquatic Botany}, \textbf{72}, 261-274.

\noindent
Vollenweider, R.A. (1975) Input-output models with special reference to the phosphorus loading concept in limnology. \textit{Schweizerland Journal of  Hydrology}, \textbf{37}, 53-84.

\noindent
Weisner, S.E.B., Strand, J.A. \& Sandsten, H. (1997) Mechanisms regulating abundance of submerged vegetation in shallow eutrophic lakes. \textit{Oecologia}, \textbf{109}, 592-599.

\noindent
Zhang, J., J\o rgensen, S.E., Tan, C.O. \& Beklioglu, M. (2003) A structurally dynamic modelling - Lake Mogan, Turkey as a case study. \textit{Ecological Modeling}, \textbf{164}, 103-120.

\clearpage

\noindent
Table 1. Average ($\pm$ SD) graph theoretical indices of the individual cognitive maps and the indices of the collective cognitive map of the scientists. 

\vspace{0.3in}

\begin{tabular}{lcc}
\hline
\textbf{Indices}&\textbf{Individual Maps}&\textbf{Collective Map}\\
Maps (n)&8&1\\
Variables (N)&$14.6 \pm 3.3$&32\\
Number of Connections (C)&$30.9 \pm 5.7$&113\\
Number of Transmitter Variables (T)&$3.8 \pm 2.5$&6\\
Number of Receiver Variables (R)&$3.4 \pm 2.1$&1\\
Number of Ordinary Variables (O)&$10.6 \pm 3.1$&25\\
\hline
Distance&$1.56 \pm 0.44$&2.79\\
C/N&$3.9 \pm 0.7$&3.5\\
Complexity (R/T)&$1.4 \pm 1.2$&0.2\\
Density (C/N(N-1))&$0.031 \pm 0.006$&0.114\\
Hierarchy Index&$0.015 \pm 0.003$&0.005\\
\hline
\end{tabular}

\pagebreak

\noindent
Table 2. Most mentioned variables listed in decreasing order of number of times occurring in the maps drawn by more than two scientists. 

\vspace{0.15in}

\begin{tabular}{lc}
\hline
\textbf{Variable}&\textbf{Times mentioned}\\
Maps (N)&8\\
Suspended Solids&8\\
Water Clarity&7\\
Zooplanktivorous Fish&7\\
Submerged Plants&7\\
Benthivorous Fish&7\\
Phosphorus Concentration&7\\
Water Level&6\\
Waves/Wind Action&6\\
Zooplankton&6\\
Herbivorous Birds&5\\
Shoreline Slope&4\\
Piscivorous Fish&4\\
Periphyton&4\\
Benthic Biomass&4\\
Benthic Production&3\\
Sediment Stability&3\\
Sediment P-Pool&3\\
Fetch&3\\
Temperature&2\\
Nitrogen Concentration&2\\
Zooplankton Grazing Rate&2\\
\hline
\end{tabular}

\pagebreak
\noindent
Table 3. The strongest connections in the collective cognitive map. The relations are given in decreasing weight. The direction of the connection is from variable 1 to variable 2.

\vspace{0.3in}

\begin{tabular}{llc}
\hline
\textbf{Variable 1}&\textbf{Variable 2}&\textbf{Weight}\\
Suspended Solids&Water Clarity&-0.84\\
Phosphorus&Phytoplankton&0.78\\
Water Clarity&Submerged Plants&0.72\\
Benthivorous Fish&Suspended Solids&0.59\\
Zooplankton&Phytoplankton&-0.47\\
Benthivorous Fish&Submerged Plants&-0.44\\
Submerged Plants&Suspended Solids&-0.44\\
\hline
\end{tabular}

\pagebreak

\noindent
\textbf{Figure Captions}

\noindent
\textbf{Figure 1.} Accumulation curve of number of new variables versus the numer of added maps. The dashed lines represent $\pm$ SD.

\noindent
\textbf{Figure 2.} Most central variables in the collective map. Bars indicate centrality, which is the sum of the indegree and outdegree of a variable showing the importance of the role of the variable. 

\noindent
\textbf{Figure 3.} The cognitive interpretation diagramm (CID) of the collective map of 8 scientists. Only the strongest connections ($\geq 0.25$) are drawn. The thickness of the lines indicates the relative strength of the connections (2 pixel corresponds to a connection with strength 0.125). Solid lines represent positive connections, dashed lines negative connections. The generic model of Scheffer et al. (1993a) is superimposed on our CID and is represented by two concentric circles. Binary connections that are also present in Scheffer et al. have a circle at the arrowheads and have all the same signs as ours.

\noindent
\textbf{Figure 4.} imulation of steady state conditions of the variables obtained from the collective cognitive map with no management option.

\noindent
\textbf{Figure 5.} The change from steady state under different management scenarios. Scenario A: Submerged plant harvest; B: Nutrient reduction; C: Fish removal without nutrient reduction; D: Biomanipulation, fish removal accompanied with  nutrient reduction.

\pagebreak

\begin{figure}[ht]
\begin{center}
\includegraphics[width=\textwidth]{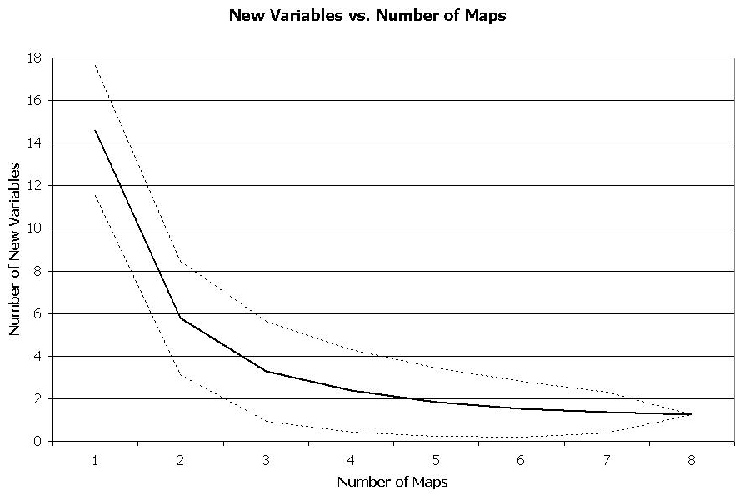}
\caption{}
\end{center}
\end{figure}

\clearpage

\begin{figure}[ht]
\begin{center}
\includegraphics[width=\textwidth]{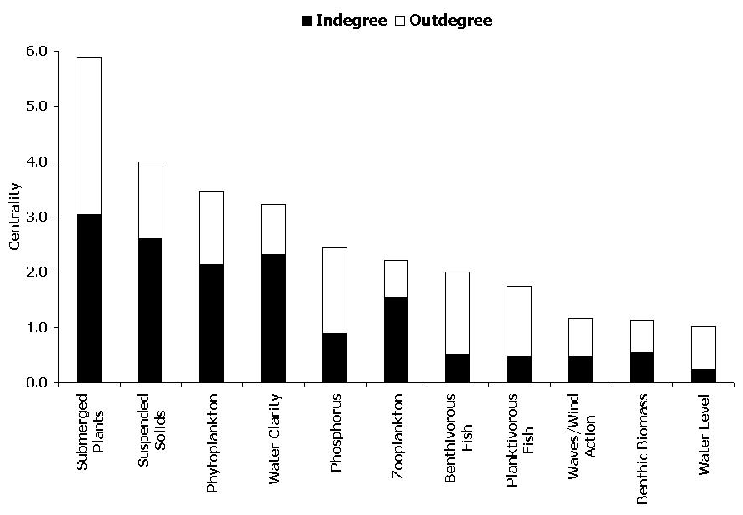}
\caption{}
\end{center}
\end{figure}

\pagebreak

\begin{figure}[ht]
\begin{center}
\includegraphics[width=\textwidth]{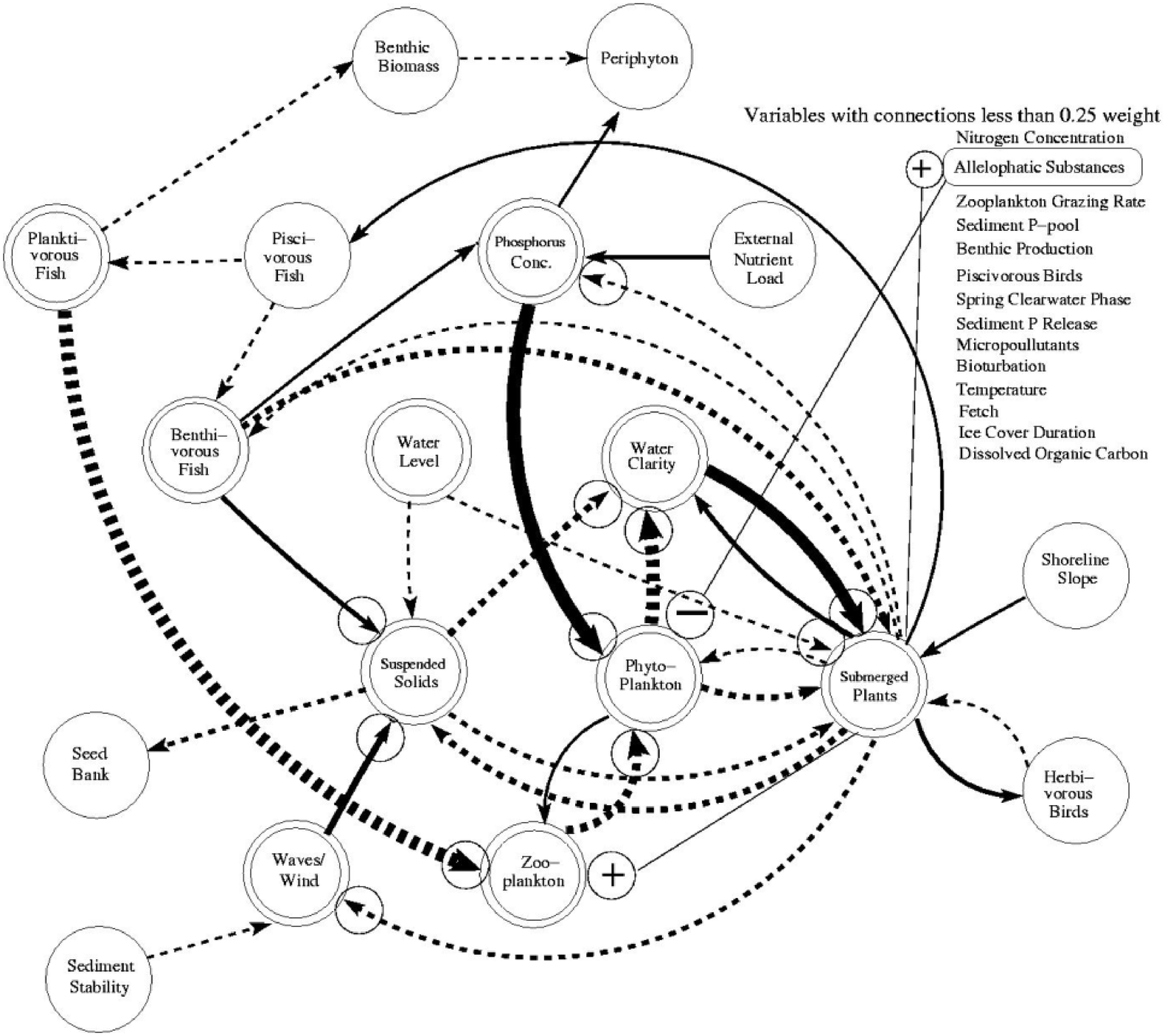}
\caption{}
\end{center}
\end{figure}

\pagebreak

\begin{figure}[ht]
\begin{center}
\includegraphics[width=\textwidth]{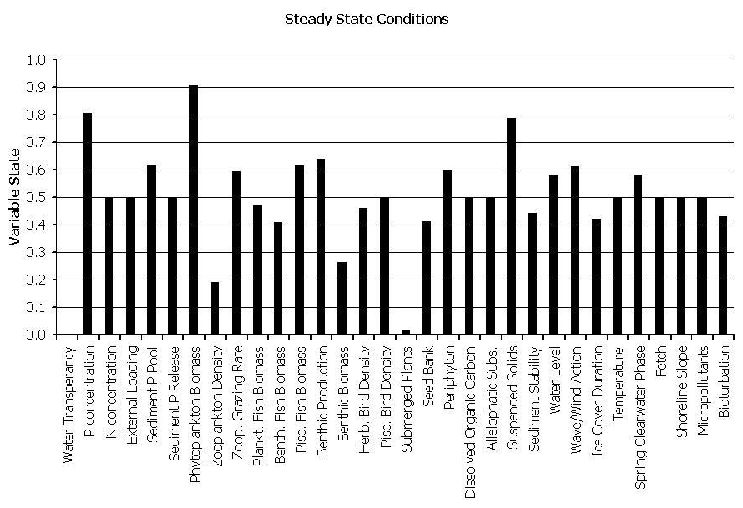}
\caption{}
\end{center}
\end{figure}

\begin{figure}[ht]
\begin{center}
\includegraphics[height=8.1in]{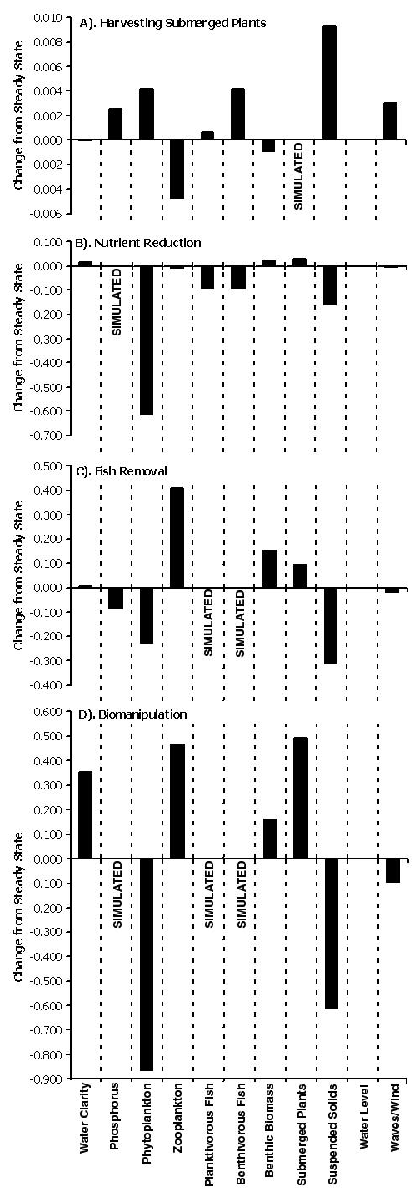}
\caption{}
\end{center}
\end{figure}

\end{document}